\documentclass[aps,twocolumn,showpacs,preprintnumbers,noshowpacs,superscriptaddress,6pt]{revtex4-2}
\usepackage[utf8]{inputenc}
\usepackage[croatian, english]{babel}
\usepackage{bm}
\usepackage{siunitx}
\usepackage{amsmath} %расширенный пакет математики
\usepackage{amsfonts}
\usepackage{diagbox}
\usepackage[usenames]{color}
\usepackage{colortbl}
\usepackage{booktabs}
\usepackage{multirow}
\usepackage[unicode, colorlinks=true, linkcolor=black, urlcolor=blue, citecolor=blue, pdftex]{hyperref}
\usepackage{color,soul}
\usepackage{ulem}

\usepackage{color, colortbl}
\definecolor{Gray}{gray}{0.9}
\definecolor{LightCyan}{rgb}{0.88,1,1}
\definecolor{BLUE}{rgb}{0.0,0.0,1.0}

\begin{document}

\title{Calculations of the binding-energy differences for highly-charged Ho and Dy ions}

\author{I.\,M.\,Savelyev}
\email[]{savelevigorm@gmail.com}
\affiliation{Department of Physics, St. Petersburg State University, 7/9 Universitetskaya nab., 199034 St. Petersburg, Russia}

\author{M.\,Y.~Kaygorodov}
\affiliation{Department of Physics, St. Petersburg State University, 7/9 Universitetskaya nab., 199034 St. Petersburg, Russia}

\author{Y.\,S.\,Kozhedub}
\affiliation{Department of Physics, St. Petersburg State University, 7/9 Universitetskaya nab., 199034 St. Petersburg, Russia}

\author{I.\,I.~Tupitsyn}
\affiliation{Department of Physics, St. Petersburg State University, 7/9 Universitetskaya nab., 199034 St. Petersburg, Russia}

\author{V.\,M.~Shabaev}
\affiliation{Department of Physics, St. Petersburg State University, 7/9 Universitetskaya nab., 199034 St. Petersburg, Russia}
\affiliation{B. P. Konstantinov Petersburg Nuclear Physics Institute of National Research Centre ``Kurchatov Institute'', Gatchina, 188300 Leningrad District, Russia}

\begin{abstract}
The binding-energy differences for~$^{163}\mathrm{Ho}^{q+}$ and~$^{163}\mathrm{Dy}^{q+}$ ions with ionization degrees~$q = 38$, $39$, and~$40$ are calculated. 
The calculations are performed using the large-scale relativistic configuration-interaction and relativistic coupled-clusters methods.
The contributions from quantum-electrodynamics, nuclear-recoil, and frequency-dependent Breit-interaction effects are taken into account. 
The final uncertainty does not exceed~$1$~eV.
Combining the obtained results with the binding-energy difference for neutral atoms calculated in [\href{https://doi.org/10.1103/PhysRevA.105.012806}{Savelyev \textit{et al.}, Phys. Rev. A \textbf{105}, 012806 (2022)}], we get the secondary differences of the ion-atom binding energies. These values can be used to evaluate the amount of energy released in the electron capture process in~$^{163}\mathrm{Ho}$ atom (the $Q$ value), provided mass differences of highly charged ions $^{163}\mathrm{Ho}^{q+}$ and~$^{163}\mathrm{Dy}^{q+}$ is known from experiment.
The $Q$ value is required by experiments on the determination of the absolute scale of the electron neutrino mass by studying the beta-decay process.

\end{abstract}

\maketitle
The neutrinos are one of the most intriguing topics of the modern physics~\cite{2020_ZuberK}. 
On the one hand, the neutrino is a massless particle within the Standard Model, on the other hand, observations of neutrino oscillations established that neutrino must have nonzero mass.
Nevertheless, the oscillation experiments are only sensitive to the absolute squared mass–differences between the neutrino mass eigenstates and do not allow one to determine the absolute mass–scale of neutrinos.
Constraints on the sum of the neutrino masses can be obtained from the analysis of the cosmological data~\cite{2017_VagnozziS_PRD96, 2020_IvanovM_PRD101}.
According to the CPT invariance of the Standard Model, the mass of the neutrino must be exactly equal to the mass of the antineutrino.
However, at present, upper limits on the electron neutrino and antineutrino masses differ by several orders of magnitude~\cite{2022_PDG}.
All these points rise undoubted interest to direct model-independent experimental determination of the neutrino mass.
\par
The best upper direct limit on the electron \textit{antineutrino} mass is obtained by KATRIN collaboration~\cite{katrin2022direct}.
The value of $0.8$~eV is obtained by performing the kinematic analysis of $\beta^{-}$~decay in tritium.
The best upper laboratory limit on the electron \textit{neutrino} mass is about two orders of magnitude larger. So, in~\cite{1987_SpringerP_PhysRevA}, using the spectroscopy of X-rays emitted in the electron-capture (EC) decay of isotope $^{163}\mathrm{Ho}$, a limit of $225$~eV was established.
In an experiment of a completely different type, based on the study of bound-state $\beta^{-}$ decay of the bare $^{163}\mathrm{Dy}$ nucleus, a limit of 410~eV was obtained~\cite{1992_JungM_PRL69}.
Several collaborations~\cite{2015_AlpertB_EPJC75, 2016_CroseM_JLTP184, 2017_GastaldoL_EurPhysJSpecTop} aim to improve the current laboratory limit on the electron neutrino mass up to a few eV and make it comparable with the value of the electron antineutrino limit.
The experiments are also based on study of the nuclear-electron-capture process in neutral $^{163}\mathrm{Ho}$ atom, but this time using a more accurate calorimetric method.
Recently, within the framework of ECHo experiment~\cite{2017_GastaldoL_EurPhysJSpecTop}, the upper limit for the electron neutrino mass has been improved to a value of about~$150$~eV~\cite{Velte2019}.
\par
In order to extract the limit on the electron neutrino mass with the eV accuracy from the experiment, one has to know in advance the mass difference between the decay isotopes $^{163}$Ho and $^{163}\mathrm{Dy}$, also called the $Q$ value, at least, with the same accuracy.
The measurement of the mass difference at the required level of accuracy can be performed for highly charged ions utilizing Penning-trap mass spectrometers~\cite{2020_RischkaA_PRL124, 2021_FilianinP_PRL127}.
Recently, the mass difference $^{163}$Ho and $^{163}$Dy ions were measured for ionization degrees $38$, $39$, and $40$~\cite{2023_Eliseev_EurPhysJA}.
\par
To convert the mass difference of
highly charged $^{163}$Ho and $^{163}$Dy ions to the mass difference of
neutral $^{163}$Ho and $^{163}$Dy atoms, which is the required $Q$ value, the binding-energy difference for the atoms and ions has to be calculated at an appropriate level of uncertainty.
In Ref.~\cite{Savelyev_2022}, the calculations of the binding-energy differences for the $^{163}\mathrm{Ho}^{q+}$ and $^{163}\mathrm{Dy}^{q+}$ ions were performed for ionization degrees $q=30$, $48$, and $56$. 
The aim of the present work is to extend the results of Ref.~\cite{Savelyev_2022} with the calculations for the ionization degrees needed for the experiment~\cite{2023_Eliseev_EurPhysJA}.
The atomic units are used throughout the paper.
\par
We consider the mass difference~$\Delta m^q$ of the ions~$^{163}$Ho$^{q+}$ and~$^{163}$Dy$^{q+}$ with the same ionization degree~$q$,
\begin{equation}
    \Delta m^q = \Delta m_n + m_e + \Delta E^q,
\end{equation}
where~$\Delta m_n$ is the mass difference of the~$^{163}$Ho and~$^{163}$Dy nuclei, $m_e$ is the electronic mass, $\Delta E^{q}$ is the difference of the total electronic binding energies of the ions.
The mass difference of the neutral~$^{163}$Ho and~$^{163}$Dy atoms corresponds to the case~$q=0$:
\begin{equation}
    \Delta m^0 = \Delta m_n + m_e + \Delta E^0.
\end{equation}
The mass difference of the neutral atoms~$\Delta m^0$ is related to the mass difference of the ions~$\Delta m^q$ by
\begin{equation}
    \Delta m^0 = \Delta m^q + \Delta E^{0,q},
\end{equation}
where the secondary difference of the binding energies,
\begin{equation}
     \Delta E^{0,q} = \Delta E^0 - \Delta E^q,
\end{equation}
is introduced.
\par
The present calculations are based on the relativistic Dirac-Coulomb (DC) and Dirac-Coulomb-Breit (DCB) Hamiltonians
\begin{align}\label{H_np}
\hat H_{\textrm{DC}}&=\Lambda^+ (\hat H_{\textrm{D}}+\hat H_{\textrm{C}}) \Lambda^+, \\
\hat H_{\textrm{DCB}}&=\Lambda^+ (\hat H_{\textrm{D}}+\hat H_{\textrm{C}}+\hat H_{\textrm{B}}) \Lambda^+, 
\end{align}
where \( \hat H_{\textrm{D}}\) is the sum of the one-electron Dirac Hamiltonians:
\begin{equation}\label{H_D}
    \hat H_{\mathrm{D}} = \sum_{i=1}^N \left[ (\bm{\alpha}_i \cdot \bm{p}_i)c+(\beta-1)mc^2+V(r_i) \right],
\end{equation}
and \(\hat H_{\textrm{C}}\) and \(\hat H_{\textrm{B}}\) are the two-electrons operators of the Coulomb and Breit interactions, respectively,
\begin{equation}
    \hat H_{\mathrm{C}} = \frac{1}{2}\sum_{i\neq j}^N \frac{1}{r_{ij}},
\end{equation}
\begin{equation}
    \hat H_{\mathrm{B}} = -\frac{1}{2} \sum_{i \neq j}^N \frac{1}{2r_{ij}}\Big[\bm{\alpha}_i\cdot\bm{\alpha}_j+\frac{(\bm{\alpha}_i\cdot\bm{r}_{ij})(\bm{\alpha}_j\cdot\bm{r}_{ij})}{r_{ij}^2}\Big].
\end{equation}
Here $\bm{\alpha}$ is a vector of the Dirac matrices, $\bm{p}$ is the momentum operator, $\bm{r}_{ij}$ is a position of the $i$-th electron relative to the $j$-th one, $r_{ij} = |\bm{r}_{ij}|$, and $N$ is the total number of electrons. 
The projector $\Lambda^+$ ensures that the Hamiltonian acts in space of states which corresponds to the positive-energy spectrum of the Dirac-Fock Hamiltonian.
\par
The main method used to calculate $\Delta E^{q}$ is the relativistic configuration interaction method in the basis of the Dirac-Fock-Sturm orbitals (CI-DFS)~\cite{2003_TupitsynI_PhysRevA, 2005_TupitsynI_PhysRevA, 2018_TupitsynI_PhysRevA}. 
To take into account the quantum-electrodynamics (QED) effects, the model-QED-operator approach~\cite{2013_ShabaevV_PhysRevA, 2015_ShabaevV_CompPhysComm, 2018_ShabaevV_CompPhysComm} is employed.
The model-QED operator $\hat{V}_{\mathrm{QED}}^{\mathrm{mod}}$ allows one to approximately account for the lowest order vacuum-polarization and self-energy corrections in many-electron systems.
The operator $\hat{V}_{\mathrm{QED}}^{\mathrm{mod}}$ is used by including it in the many-electron DCB Hamiltonian 
\begin{equation}
    \hat H_{\mathrm{DCBQ}} =\Lambda^+ (\hat H_{\mathrm{D}}+\hat H_{\mathrm{C}}+\hat H_{\mathrm{B}} + \hat{V}_{\mathrm{QED}}^{\mathrm{mod}}) \Lambda^+ .
\end{equation}
\par
The frequency-dependent Breit-interaction correction is evaluated as the expectation value of the frequency dependent part of the one-photon-exchange operator considered in the Coulomb gauge with the CI-DFS many-electron wave function. 
The nuclear-recoil effect is evaluated as the expectation value of the relativistic nuclear-recoil Hamiltonian~\cite{Shabaev1985TMP_Recoil, Shabaev1988, Palmer1987, 1998_ShabaevV_PhysRevA} with the many-electron wave function constructed by CI-DFS.
\par
To verify the accuracy of the CI-DFS correlation treatment within the DC Hamiltonian framework, we performed the calculations using another approach, namely the relativistic single-reference coupled-clusters method involving fully iterative single, double and perturbatively triple cluster amplitudes (CCSD(T)).
The software package DIRAC23~\cite{2020_SaueT_JCP152, bast_r_2023_7670749} is used for this purpose.
\par
The Fermi model of the nuclear charge distribution is used in our CI-DFS calculations, whereas in the CCSD(T) calculations, the Gaussian model is employed, but for the properties under consideration the difference is negligible.
The values of the root-mean-square (RMS) radii are taken from Ref.~\cite{Ang13}.

\par
We approach the problem of calculating $\Delta E^q$ by progressive stages.
At the first stage, we evaluate the total binding-energy differences for Ho and Dy ions at the DC Hamiltonian level by means of the CI-DFS method.
The ground-state configurations of the considered Ho and Dy ions with $q=38$, $39$, and $40$ are given in Table~\ref{tab:0}.
\begin{table}[htbp]
    \centering
    \caption{The ground-state configurations of Ho$^{q+}$ and Dy$^{q+}$ ions.}
\begin{tabular}{
l
l
l
l
l
}
\toprule
\multirow{2}{*}{$q$} & \multicolumn{2}{c}{Ho$^{q+}$} & \multicolumn{2}{c}{Dy$^{q+}$}              \\
         & Configuration                        & $J$    & Configuration        & $J$                 \\
\midrule
$38$     & $[\mathrm{Ar}] 3d^{10}4s^1$ & $1/2$  & $[\mathrm{Ar}] 3d^{10}$ & $0$     \\
$39$     & $[\mathrm{Ar}] 3d^{10}$     & $0$    & $[\mathrm{Ar}] 3d^{9}$  & $5/2$   \\
$40$     & $[\mathrm{Ar}] 3d^{9}$      & $5/2$  & $[\mathrm{Ar}] 3d^{8}$  & $4$     \\    
\bottomrule
% & ${6d^{8}7s^2\,\,J=4}$ & ${6d^{10}7s^2}$ \\     

    % $q$& \multicolumn{2}{c}{$\textrm{Ho}^{q+}$}   &  \multicolumn{2}{c}{$\textrm{Dy}^{q+}$} \\ 
    %   \midrule
    % $38$&$[\mathrm{Ar}] 3d^{10}4s^1$& 0.5&$[\mathrm{Ar}] 3d^{10}$& 4\\
    %                 $39$&$[\mathrm{Ar}] 3d^{10}$& 0&$[\mathrm{Ar}] 3d^9$& 2.5\\
    %   $40$ &$[\mathrm{Ar}] 3d^9$& 2.5 &  $[\mathrm{Ar}] 3d^8$& 4 \\
    %   \bottomrule
    \end{tabular}
    \label{tab:0}
\end{table}
In the CI-DFS calculations, all the occupied orbitals are divided into the valence and core ones. 
The $1s2s2p$ orbitals are assigned to the frozen core, the remaining occupied orbitals are active.
The single and double excitations from the active orbitals to the virtual ones are considered.
The convergence of $\Delta E^q$ with respect to the number of virtual orbitals is analyzed. 
We start with one virtual $s$ orbital and expand the set of the virtual orbitals in two directions: by successively adding new virtual orbitals with larger principal quantum number $n$ and with larger orbital quantum number $l$.
Our calculations include up to five virtual orbitals for each orbital quantum number $l$ up to $i$.
The final results are obtained by extrapolating the values to the infinite basis-set limit.
\par
At the next stage, we evaluate the contribution from the frozen core.
To this end, we perform the CI-DFS calculations using smaller number of virtual orbitals adapted to correlate strongly bound $1s2s2p$ electrons.
\par
Further, we verify the correctness of our CI-DFS results by performing the CCSD(T) calculations. 
Within the CCSD(T) approach, all electrons are correlated and the standard basis sets~\texttt{dyall.ae3z} and~\texttt{dyall.ae4z} are employed.
The convergence of the CCSD(T) results is studied in the same manner as in the CI-DFS scheme, with the extrapolation to the infinite basis-set limit being used. It should be noted that the contribution of the perturbative triple
cluster amplitude is negligible for the considered systems.
\par
The differences of the ground-state energies for $^{163}$Ho$^{q+}$ and $^{163}$Dy$^{q+}$ ions with $q=38$, $39$, and $40$ evaluated by means of the CI-DFS and CCSD(T) methods using the DC Hamiltonian are presented in Table~\ref{tab:1}.
\begin{table}[htbp]
\centering
\caption{The ground-state energy differences for $^{163}$Ho$^{q+}$ and $^{163}$Dy$^{q+}$ ions obtained using the CI-DFS and CCSD(T) methods with the DC Hamiltonian~(a.u.).}
\begin{tabular}{
l 
S[table-format=-3.4(1)] 
S[table-format=-3.4(1)] 
}
\toprule
\multicolumn{1}{c}{$q$} & \multicolumn{1}{c}{CI-DFS} & \multicolumn{1}{c}{CCSD(T)} \\ 
\midrule
38 &  -461.298(7)  &  -461.305(11) \\
39 &  -502.088(16) &  -502.089(16) \\
40 &  -500.931(17) &  -500.934(17) \\
\bottomrule
\end{tabular}
\label{tab:1}
\end{table}  
The results obtained using two conceptually different methods agree within the estimated uncertainties. 
\par

To evaluate the Breit, frequency-dependent Breit, QED, and nuclear-recoil corrections, a series of the CI-DFS calculations with active~$3s3p3d$ (and~$4s$ for $\textrm{Ho}^{38+}$) orbitals is carried out. It turns out that the sufficient accuracy for these corrections can be achieved using a smaller basis set than for the DC results.
The Breit and QED corrections are obtained using the Hamiltonians $\hat H_{\mathrm{DCB}}$ and $\hat H_{\mathrm{DCBQ}}$, respectively, whereas the frequency-dependent part of the Breit-interaction correction and the nuclear-recoil correction are calculated as expectation values of the corresponding operators.
The errors associated with the choice of the nuclear model and with the RMS radii are estimated by considering different charge distributions and varying the RMS radii within their experimental uncertainties.

\par
The final results for the total binding-energy differences for~$^{163}$Ho$^{q+}$ and~$^{163}$Dy$^{q+}$ ions for various degrees of ionization~$q=38$, $39$, and $40$, $\Delta E^{q}$, as well as the additional contributions from the DCB Hamiltonian, $\Delta E^{q}_{\mathrm{DCB}}$, the QED corrections, $\Delta E^{q}_{\mathrm{QED}}$, and the frequency-dependent Breit-interaction correction,~$\Delta E^{q}_{\mathrm{BRFD}}$, are presented in Table~\ref{tab:2}. The nuclear-recoil correction is found to be about $0.001$~a.u. for all considered $q$. This correction is included in $\Delta E^{q}_{\mathrm{DCB}}$.
\begin{table}[htbp]
\centering
\caption{
Contributions to the ground-state energy differences for $^{163}$Ho$^{q+}$ and $^{163}$Dy$^{q+}$ ions,~$\Delta E^q$, from the electron-correlation effects within the Breit approximation,~$\Delta E^q_{\mathrm{DCB}}$, from the frequency-dependent Breit interaction,~$\Delta E^q_{\mathrm{BRFD}}$, and from QED,~$\Delta E^q_{\mathrm{QED}}$ (a.u.).
In the last column, the total ground-state energy differences are presented. The numbers in the first parentheses show numerical accuracy, whereas the numbers in the second parentheses represent the uncertainty associated with the finite nuclear size.}
\begin{tabular}{
l 
S[table-format=-4.4] 
S[table-format=-1.5]
S[table-format=1.4] 
S[table-format=-3.4(4), parse-numbers=false]
}
\toprule
\multicolumn{1}{c}{$q$}  &
\multicolumn{1}{c}{$\Delta E^{q}_{\mathrm{DCB}}$}  &
\multicolumn{1}{c}{$\Delta E^{q}_{\mathrm{BRFD}}$}  &
\multicolumn{1}{c}{$\Delta E^{q}_{\mathrm{QED}}$}  &
\multicolumn{1}{c}{$\Delta E^{q}$}  \\ 
\midrule
38 & -460.678  & -0.0140  &  0.439  &-460.253(9)(13) \\
39 & -501.446  & -0.0146  &  0.405  &-501.056(16)(13) \\
40 & -500.291  & -0.0146  &  0.405  &-499.901(17)(13) \\
\bottomrule
\end{tabular}
\label{tab:2}
\end{table}  
The values~$\Delta E^{q}$ have two sources of uncertainty: in the first parentheses, the uncertainty due to the correlation treatment within the CI-DFS method is shown, meanwhile, in the the second parentheses, the uncertainty due to the nuclear parameters is presented.
The evaluated energy differences for the ions,~$\Delta E^{q}$, have smaller uncertainty associated with the electronic correlation than the energy difference for the neutral atoms from Ref.~\cite{Savelyev_2022} due to simpler electronic structure of the ions. Note that our results obtained by the Dirac-Fock method without taking into account the correlation effects are consistent  with the results of \cite{2004_RodriguesG_ADNDT86}.
\par
Combining the calculated energy differences for the ions,~$\Delta E^{q}$, with the energy difference for the neutral atoms,~$\Delta E^{0}$, evaluated in Ref.~\cite{Savelyev_2022}, we obtain the secondary differences of the ion-atom binding energies,~$\Delta E^{0,q}$, which are presented in units of eV in Table~\ref{tab:3}.
\begin{table}[htbp]
\centering

\caption{The contributions to the %binding-energy differences of $q$ ionized electrons in
secondary differences of the ion-atom binding energies for
$^{163}\mathrm{Ho}$ and $^{163}\mathrm{Dy}$ atoms,~$\Delta E^{0,q}$, from the electronic correlations within the Breit approximation,~$\Delta E^{0,q}_{\mathrm{DCB}}$, the frequency-dependent part of the Breit interaction,~$\Delta E^{0,q}_{\mathrm{BRFD}}$, and the QED effects,~$\Delta E^{0,q}_{\mathrm{QED}}$~(eV). The total secondary differences,~$\Delta E^{0,q}$, are given in the last column.}
\label{tab:3}
\begin{tabular}{l 
S[table-format=4.1] 
S[table-format=-1.2] 
S[table-format=-1.3] 
S[table-format=4.1(1)]
}

\toprule
\multicolumn{1}{c}{$q$}  &
\multicolumn{1}{c}{$\Delta E^{0,q}_{\mathrm{DCB}}$}  &
\multicolumn{1}{c}{$\Delta E^{0,q}_{\mathrm{BRFD}}$}  &
\multicolumn{1}{c}{$\Delta E^{0,q}_{\mathrm{QED}}$}  &
\multicolumn{1}{c}{$\Delta E^{0,q}$}  \\
\midrule
38  &   37.7 & -0.03  & -0.77  & 36.9(7)   \\
39  & 1147.0 & -0.01  &  0.17  & 1147.2(8) \\
40  & 1115.6 & -0.01  &  0.18  & 1115.8(8) \\
\bottomrule

\end{tabular}
\end{table}
The uncertainties associated with the finite nuclear-size effects are canceled in the secondary differences~$\Delta E^{0,q}$ and the uncertainties of the final results are mainly determined by the binding-energy difference error for the neutral atoms.

\par
In conclusion, the total binding-energy differences for $^{163}\mathrm{Ho}^{q+}$ and $^{163}\mathrm{Dy}^{q+}$ ions for degrees of ionization $q=38$, $39$, and $40$ are calculated by means of the CI-DFS method with sub-eV uncertainty. 
The electronic correlations are evaluated within the framework of the DCB Hamiltonian.
The QED, frequency-dependent Breit-interaction, and nuclear-recoil corrections are taken into account.
Combining the calculated energies differences for ions with the energies difference for neutral atoms reported in Ref.~\cite{Savelyev_2022}, we obtain the secondary differences of the ion-atom binding energies.
These results can be used to recalculate the mass difference between ions $^{163}\mathrm{Ho}^{q+}$ and $^{163}\mathrm{Dy}^{q+}$ to the $Q$ value of the electron capture process in $^{163}\mathrm{Ho}$ atom, which is required for upcoming experiments on improving the upper limit of the electron neutrino mass.
\par
This work was supported by Russian Science Foundation grant No.~22-62-00004.

\bibliographystyle{my-h-physrev.bst}
\bibliography{main}
\end{document}